\documentclass[preprint,12pt]{elsarticle}

\usepackage{times}
\usepackage{soul}
\usepackage{url}
\usepackage[hidelinks]{hyperref}
\usepackage[utf8]{inputenc}
\usepackage[small]{caption}
\usepackage{graphicx} 
\usepackage{amsmath}
\usepackage{amsthm}
\usepackage{amssymb}
\usepackage{booktabs}
\usepackage{algorithm}
\usepackage{algorithmic} 
\usepackage[switch]{lineno}
\usepackage{multirow}
\usepackage{float}
\usepackage{subfigure}
\usepackage{graphicx}
\usepackage[export]{adjustbox}
\usepackage{xcolor}
\usepackage{amsmath} 
\usepackage{bm}      
\usepackage {verbatim}

\usepackage{xspace}
\makeatletter
\DeclareRobustCommand\onedot{\futurelet\@let@token\@onedot}
\def\@onedot{\ifx\@let@token.\else.\null\fi\xspace}
\def\eg{\emph{e.g}\onedot} 
\def\ie{\emph{i.e}\onedot}

\makeatother

\nolinenumbers
\journal{Expert Systems With Applications}

\begin{document}

\begin{frontmatter}

\title{Enhancing Traffic Signal Control through Model-based Reinforcement Learning and Policy Reuse}

\author[label1]{Yihong Li}
\author[label1]{Chengwei Zhang\corref{cor1}}
\author[label1]{Furui Zhan\corref{cor1}}
\author[label1]{Wanting Liu}
\author[label1]{Kailing Zhou}
\author[label1]{Longji Zheng}

\affiliation[label1]{organization={Dalian Maritime University},
            addressline={1 Linghai Road}, 
            city={Dalian},
            postcode={116026}, 
            state={Liaoning},
            country={China}}

\cortext[cor1]{Corresponding authors}
\cortext[cor2]{E-mail addresses: liyihong@dlmu.edu.cn (Y. Li), chenvy@dlmu.edu.cn (C. Zhang), izfree@dlmu.edu.cn (F. Zhan), wantyluu@dlmu.edu.cn (W. Liu), zhoukl@dlmu.edu.cn (K. Zhou), zhenglongji@dlmu.edu.cn (L. Zheng).}




\begin{abstract}
Multi-agent reinforcement learning (MARL) has shown significant potential in traffic signal control (TSC). However, current MARL-based methods often suffer from insufficient generalization due to the fixed traffic patterns and road network conditions used during training. This limitation results in poor adaptability to new traffic scenarios, leading to high retraining costs and complex deployment. To address this challenge, we propose two algorithms: PLight and PRLight. PLight employs a model-based reinforcement learning approach, pretraining control policies and environment models using predefined source-domain traffic scenarios. The environment model predicts the state transitions, which facilitates the comparison of environmental features. PRLight further enhances adaptability by adaptively selecting pre-trained PLight agents based on the similarity between the source and target domains to accelerate the learning process in the target domain. We evaluated the algorithms through two transfer settings: (1) adaptability to different traffic scenarios within the same road network, and (2) generalization across different road networks. The results show that PRLight significantly reduces the adaptation time compared to learning from scratch in new TSC scenarios, achieving optimal performance using similarities between available and target scenarios.

\end{abstract}


\end{frontmatter}

\section{Introduction}
\label{sec:1}
As urbanization progresses, traffic congestion becomes increasingly severe, significantly affecting people's daily lives. Traffic Signal Control (TSC) systems are considered effective solutions to alleviate congestion by optimizing traffic light timings and improving road efficiency~\cite{haydari2020deep}. Recently, Multi-Agent Reinforcement Learning (MARL) methods have gained widespread attention for TSC tasks. However, MARL still faces many challenges when dealing with complex and variable traffic scenarios. Most existing methods focus on training models under predefined demand scenarios for specific times of day (TOD)~\cite{yoon2021transferable}. The partial observability of the environment or sparse reward feedback requires even well-trained agents to perform extensive exploration when encountering unfamiliar TOD scenarios, thereby limiting their ability to utilize environmental knowledge.

As a result, it is difficult to make efficient decisions and retrain the model in new environments, which wastes computational resources and makes real-world deployment challenging. Transfer learning (TL)~\cite{zhu2023transfer} is commonly used to address issues such as low sampling efficiency, slow training speed, and poor generalization in new environments for MARL-based TSC methods. TL involves applying knowledge from existing tasks or domains (the source domain) to related new tasks (the target domain). The core idea is to leverage knowledge, features, or models from the source domain to assist with tasks in the target domain, especially when data are limited or model training is challenging. In adaptive TSC based on reinforcement learning, transfer learning remains an emerging topic facing significant challenges. This complexity stems from the need to transfer knowledge within the framework of Markov decision processes, which requires defining formal rules for traffic scenario models and designing mechanisms for knowledge storage and transfer to incorporate expert insights effectively.

When applying TL to practical ATSC problems, three issues need to be considered: (1) How to characterize transferable source knowledge? (2) What criteria determine the most suitable source domain knowledge? and (3) How to effectively utilize the source domain knowledge? Existing methods offer valuable insights, \eg, leverages graph neural networks to learn spatial relationships at intersections~\cite{yoon2021transferable}, transfers experience among agents within a road network through an experience classifier~\cite{norouzi2021experience}, and use meta-RL to train a generalized meta-learner for multiple tasks~\cite{zhang2020generalight,zang2020metalight}. However, these methods still face limitations, such as being restricted to single-intersection tasks, lacking cross-domain transferability, and potentially losing task-specific information. These limitations collectively highlight the need for a method that can handle multi-intersection tasks, adapt to diverse road configurations, and enable effective cross-domain transfer. To address these challenges, our method introduces an agent pool mechanism to assist the target agent in learning and adapting to new environments, by supporting knowledge transfer across multi-intersection networks and different road settings, overcoming the limitations of single-intersection tasks, network-specific solutions, and generalized models that lose task-specific details.

We identified that transferable source domain knowledge can be derived from the inherent similarities in traffic network states, as demonstrated by prior work~\cite{van2021multi} leveraging the symmetry of road topology and traffic flow characteristics. This insight suggests that experiences and decisions from similar environments can be applied to unknown traffic scenarios. Thus, our approach focuses on learning environmental models for traffic networks, comparing the similarity of states in unknown scenarios, and using policies from highly similar environments to aid decision-making in target tasks. To achieve this, we designed agent models for both source and target domains to extract and transfer knowledge. Inspired by model-based reinforcement learning approaches like MQL~\cite{oh2021model}, our network model consists of three components: an Encoder, a Decoder, and a Q-network. The Encoder extracts key features from raw observations by filtering out redundant information. The Decoder, sharing the feature extraction module with the Q-network, predicts the next observation based on the extracted features and actions, learning the state transition model. The Q-network, as the decision-making module, evaluates each action in the current state and selects the best one. Together, the Encoder-Decoder serves as the environmental model, providing external knowledge to facilitate learning in the target task.

After determining the agent model, we focus on reusing source domain knowledge for agent transfer. In the field of RL, TL can be categorized into objective transfer (reshaping the reward function), parameter transfer (reusing network weights), and behavior transfer (leveraging expert policies). Our work adopts behavior transfer~\cite{scerri2001adjustable}, which directly utilize the source task's policy to construct the target policy. This approach enables faster adaptation by utilizing high-quality experiences without additional training or network adjustments, thereby reducing computational costs and improving efficiency in ATSC tasks. Specifically, the target agent selects appropriate policies from an agent pool based on similarity weights during interactions with the environment. These weights measure the similarity between source and target tasks using the Euclidean distance between the predicted feature vectors of each policy's decoder and the actual observed feature vectors in the target domain. A smaller distance indicates higher similarity and weight. To validate our method, we conducted transfer experiments from two perspectives: adapting to varying traffic flows within the same network and generalizing across different networks. Additionally, we performed ablation studies to analyze the impact of jointly learning the environmental model and policy on overall performance. The results show that our method outperforms others in terms of performance, convergence speed, and stability.

In summary, our contributions are as follows:
\begin{itemize}
    \item We propose a model-based RL approach PLight (Predictive Light) that simultaneously learns the state transition model of task environments and their policies. Those environmental prediction models serve as a knowledge aid for learning new tasks.
    
    \item Based on the pre-trained source models, we propose a transfer RL framework PRLight (Policy-Reuse Light) using policy reuse. It calculates similarity weights between tasks to select the most appropriate policy for the target agent. This approach helps the agent explore more effectively in the early stages, thereby improving learning efficiency and training speed.

    \item We designed a diverse set of traffic network environments and evaluated our transfer method using three types of road networks and eight traffic flow datasets. We also tested the pre-trained model on all tasks to determine whether simultaneously learning the environment model and policy impacts the policy's performance.

\end{itemize}
The remainder of this paper is organized as follows. Section~\ref{sec:2} discusses related work. Section~\ref{sec:3} introduces the background knowledge of ATSC and TRL. Section~\ref{sec:4} presents the proposed algorithm. Section~\ref{sec:5} details the experiments and performance analysis. Finally, Section~\ref{sec:6} concludes the article.

\section{Related works}
\label{sec:2}

In this section, we provide a comprehensive review of the literature, focusing on three key aspects: the application of DRL in ATSC, an overview of Transfer RL and  the generalization of DRL in ATSC.

\subsection{Application of DRL in ATSC}

Research on ATSC using DRL has garnered significant attention in recent years. TSC is essentially a multi-agent cooperative task that aims to improve control efficiency, reduce congestion, and lower vehicle emissions. Effective communication between agents is key. Studies such as those by Chu et al.~\cite{chu2019multi}, Wei et al.~\cite{wei2019colight}, and Wang et al.~\cite{wang2022meta} have improved interagent communication by incorporating information or policies from neighboring agents using Graph Attention Networks (GAT) or Long Short-Term Memory (LSTM). These methods optimize system performance and form globally optimal policies through collaborative decision-making. Beyond efficiency, TSC also focuses on fairness and safety. Fairness involves equitable resource allocation among vehicles or intersections. For example, FIT~\cite{chen2021fit} balanced efficiency and fairness by modifying the reward function, inspired by the Proportional Fairness (PF) algorithm. VF-MAPPO~\cite{Liu10907776vfmappo} further optimized traffic efficiency with vehicle-level fairness by modeling fairness as a constraint. In competitive scenarios, EMVLight~\cite{su2023emvlight} prioritized emergency vehicles by modeling emergency lane formation based on the capacity of the road segment and the count of vehicles. SafeLight~\cite{du2023safelight} integrating road safety standards into DRL to achieve zero-collision intersections. To address real-world complexity, studies such as FRAP~\cite{zheng2019learning} proposed structure-agnostic models for diverse intersections, while IG-RL~\cite{devailly2021ig} used Graph Neural Networks (GNNs) to handle varying lane numbers and topologies.

Despite progress, current DRL applications in TSC face challenges. Most studies rely on trial-and-error in simulators in time-of-day (TOD) scenarios~\cite{yoon2021transferable}, resulting in low training efficiency and slow convergence. These methods often require retraining for new environments, which is impractical in real-world settings. Future research should focus on improving the generalization of DRL algorithms to enable rapid adaptation to new environments.

\subsection{Deep Transfer Reinforcement Learning}
In TSC problems, the real-world traffic conditions are dynamic and constantly changing. Most existing methods require continuous interaction with the environment through trial-and-error learning, which incurs significant costs and is impractical for real-world deployment. Therefore, enhancing the generalizability of reinforcement learning agents in TSC tasks is crucial. Transfer learning (TF) is an effective way to improve the generalization of RL, with applications broadly categorized as objective transfer, parameter transfer, and behavior transfer. Objective transfer reshapes the objective function using external knowledge, as seen in methods like PBRS~\cite{ng1999policy}, PBA~\cite{wiewiora2003principled}, DPB~\cite{devlin2012dynamic}, and DPBA~\cite{harutyunyan2015expressing}. Parameter transfer initializes the neural network of the target task with parameters from a source task, exemplified by the shared representation extraction network proposed by Rusu et al.~\cite{rusu2016progressive}. However, these methods may fail to capture environment-specific nuances. Behavior transfer, on the other hand, leverages source policy experiences to enhance exploration in the target task by selecting optimal source policies for the rapid acquisition of high-value experience samples. This approach, explored in studies such as~\cite{fernandez2006probabilistic, li2018optimal, gimelfarb2021contextual}, enables efficient exploration and accelerated learning.

Our work focuses on behavior transfer, which provides an initial optimal policy that allows the agent to adapt to the environment more quickly. Unlike objective and parameter transfer, behavior transfer does not require distillation or additional adaptation training, thus avoiding computational waste and offering a more efficient solution for DRL optimization.

\subsection{Generalization of DRL in ATSC}
Indeed, there are numerous TL works applied in ATSC tasks. TL-GNN~\cite{yoon2021transferable} addresses the limited exploration issue in predefined TOD scenarios by representing the state as a graph structure and using GNNs, enabling agents to learn spatial relationships between intersections and transfer relational knowledge to unseen data. However, it focuses on single intersections and is not applicable to multi-intersection road networks. Exp-TL-TSC~\cite{norouzi2021experience} introduces transfer learning by categorizing experiences using classifiers, providing high-quality initial experiences to aid in the learning of the target agent. GeneralLight~\cite{zhang2020generalight} and MetaLight~\cite{zang2020metalight} employ model-agnostic meta-reinforcement learning to train a meta-learner with robust initial parameters, allowing rapid adaptation to new environments with minimal fine-tuning. Similarly, Multi-View Encoder TSC~\cite{ge2021multi} divides training into three stages: specialized training for individual intersections, collaborative training of a seed apprentice, and adaptive training at target intersections. However, these methods are based on shared network parameters between environments, which can cause a loss of personalized information learned in other contexts.

Compared to these works, our method supports knowledge transfer across different multi-intersection traffic networks and leverages diverse policies from an agent pool to enable the target agent to learn tasks in the target domain more effectively.

\section{Preliminaries}
\label{sec:3}
This section provides an overview of problem modeling for the ATSC task and the configuration of the source and target domains in TL. We list the main notations in Table~\ref{tab:3-1}.

\begin{table}[ht]
\centering 
\caption{Key Notations and Definitions}
\label{tab:3-1}
\begin{tabular}{l|l}
\hline
Notation                        & Definition  \\ \hline
$o_i$                         & Observation of agent $i$ \\ \hline
$a_i$                         & Action of agent $i$  \\ \hline
$u_{i,l}$                     & Queue length of intersection $i$ \\ \hline
$r_i$                         & Individual reward of intersection $i$ \\ \hline
$\bm{\mathcal{M}}_s$            & Set of source domain tasks \\ \hline
$\mathcal{M}_s$                 & Source domain task \\ \hline
$\mathcal{M}_{tar}$             & Target domain task \\ \hline
$\mathcal{K}_s$                 & External knowledge from the source domain \\ \hline
$\mathcal{K}_{tar}$             & Internal knowledge from the target domain \\ \hline
\end{tabular}
\end{table}

\subsection{Multi-intersection ATSC as a Markov Game}
The task of TSC within a road network can be regarded as a multi-agent problem, and it is modeled using a partially observable Markov game (POMG). It can be formalized as an $ \mathcal{M}= \langle \mathcal{N},\mathcal{S},\mathcal{O},\mathcal{A},\mathcal{P},\mathcal{R},\gamma \rangle$ tuple. Here, $\mathcal{N}$ denotes a set of $n$ agents. In our setup, the number of agents $n$ is equal to the number of intersections, which means that each intersection is assigned an agent for control. $\mathcal{S}$ denotes a finite state space, and $\mathcal{O}$ denotes the observation space. Agents obtain local observations $o \in \mathcal{O}$ derived from the global state $s \in \mathcal{S}$. $\mathcal{P}$ denotes the transition probability function. The agent $i$ executes an action $a_i$ based on its local observation $o_i$. The decisions of all agents form a joint action $\bm{a} = \{a_1, a_2, ..., a_n\}$, which governs the transition of the state from $s$ to $s'$ according to the distribution $\mathcal{P}(s' \mid s, \bm{a})$. Upon transition to the next state, each intersection receives a corresponding reward. The rewards are obtained from the reward function $S \times  A_1 \times  A_2 \times  ...\times  A_n \to \mathbb{R}$. Here, $\gamma \in [0,1)$ is the discount factor.

\begin{figure}[h]
    \centering
    \includegraphics[width=0.55\textwidth]{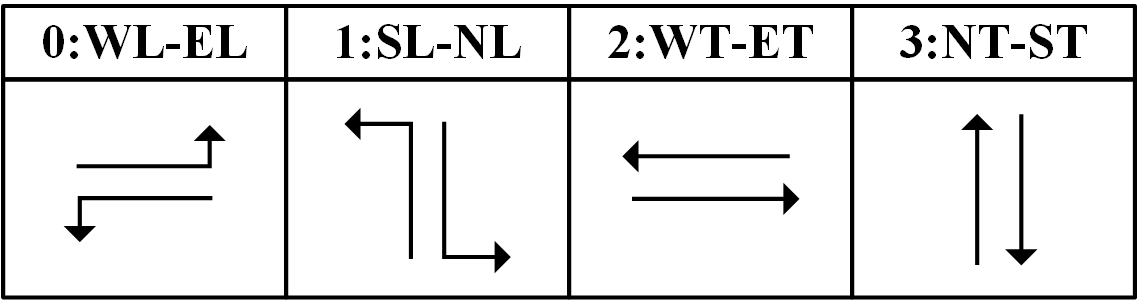}
    \caption{Four Traffic Signal Phases.}
    \label{fig:phase}
\end{figure}
In the ATSC setting, Agent $i \in \mathcal{N}$ can acquire its exclusive local observation $o_i$ from the global state, including the current intersection phase, the number of waiting vehicles, and information about its neighbors. $\mathcal{A}$ denotes the joint action space, where agent $i$ selects an action $a_i \in A_i = {0, 1, 2, 3}$ corresponding to different phases of traffic signals. The action space consists of four phases, as illustrated in Fig.~\ref{fig:phase}. The reward is defined as the length of the negative queue of the incoming lanes at each intersection to improve traffic efficiency.
\[
r_i = -\textstyle\sum_{l}u_{i,l},
\]
where $u_{i,l}$ denotes the queue length of lane $l$. For the aforementioned problem model $\mathcal{M}$, our objective is to learn an optimal policy $\pi ^*_\mathcal{M}$ with the optimal value and Q function.

\subsection{Source and Target Domains}
In transfer learning, the initial task and its associated dataset are termed the source domain $\bm{\mathcal{M}}_s = \{\mathcal{M}_1,...,\mathcal{M}_n \}$, while the task to be learned is termed the target domain $\mathcal{M}_{tar}$. The source and target domains are similar but not identical. In reinforcement learning, the task domain is typically modeled as a Markov decision process (MDP). Differences between the source and target domains can occur in any part of the MDP~\cite{zhu2023transfer}. Taking TSC as an example, differences in state space may reflect variations in the network topology and traffic distribution. Different intersection structures may have different phases, leading to differences in the action space. Domain reward definitions can also differ, resulting in variations in the reward function. Due to these factors, identical state-action pairs in both domains may have different state transition probabilities. To simplify the problem and facilitate the application of our method, we define the source and target domains in terms of their respective POMG. Although there exist certain similarities between these domains, notable differences are also present. There are certain similarities and differences between the domains:

\begin{itemize}
    \item $\mathcal{N}$: When the network structures of the tasks differ, the number of intersections may vary, leading to differences in the number of agents set up for intelligent control.
    \item $\mathcal{S}$: Differences in network structures result in variations in vehicle distributions, leading to differences in the state space of the tasks.
    \item $\mathcal{O}$: Due to the differences in the state space, the partially observable information obtained will also vary.
    \item $\mathcal{A}$: All tasks use the same action space consisting of four mutually exclusive phase actions: east-west straight, north-south straight, east-west left turn, and north-south left turn.
    \item $\mathcal{P}$: Due to the differences in the state space, the state transitions will vary.
    \item $\mathcal{R}$: The reward function is derived from the state space, hence it differs between tasks. However, to simplify the problem, the reward function is defined similarly.
    \item $\gamma$: To achieve consistent cumulative rewards, the discount factor is designed to be the same.
\end{itemize}

The similarity and difference criteria are consistent with most road network settings. Variations between tasks arise primarily from differences in state space and transition probabilities. Our approach seeks to capture and encode these differences as transferable knowledge to assist the target domain task in learning the optimal policy $\pi ^*$.
\[
\pi^* = \arg \max _{\pi}\mathbb{E}_{s\sim\rho ^{tar}_0,a\sim\pi}[Q^{\pi}_{\mathcal{M}_{tar}}(s,a)],
\]
where $\pi = \phi(\mathcal{K}_s \sim \bm{\mathcal{M}}_s,  \mathcal{K}_{tar} \sim \mathcal{M}_{tar}): \mathcal{S}^{tar} \to \mathcal{A}^{tar}$ denotes the policy for the target domain $\mathcal{M}_{tar}$ based on external knowledge from the source domain $\mathcal{K}_s$ and internal knowledge from the target domain $ \mathcal{K}_{tar}$.

\section{PreditiveLight}
This section first introduces the overall framework of PreditiveLight and then introduces each component of PreditiveLight in detail.

\subsection{Framework Overview}

\label{sec:4}
\begin{figure}[h]
    \centering
    \includegraphics[width=\linewidth]{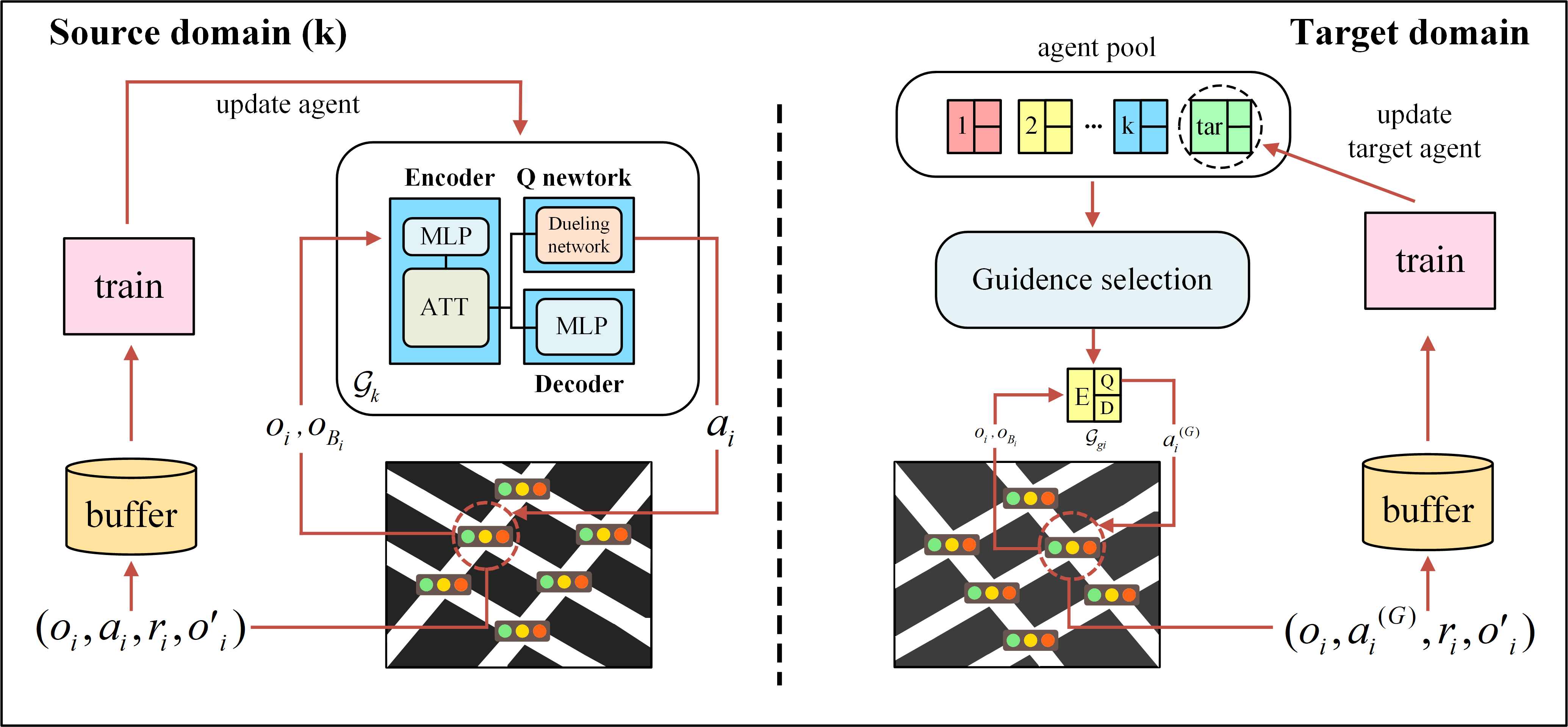} 
    \caption{Overall Architecture of the methods. The methods divided into two stages: source domain training and target domain transfer. On the left side (PLight: pre-training), the agent model’s network structure and training process for source domain tasks (illustrated using domain $k$—with identical processes for other domains) are shown, and the trained models are stored in an agent pool. On the right side (PRLight: transfer) displays the transfer and training process in the target domain task. Each block in the agent pool denotes an agent, where \textbf{E} stands for the Encoder structure, \textbf{D} for the Decoder structure, and \textbf{Q} for the Q-network.}
    \label{fig:overall}
\end{figure}

To address the challenge of using existing knowledge to support new tasks, we propose PLight and PRLight. As shown in Fig.~\ref{fig:overall}, the overall framework consists of two parts: (a) pre-training of Source Domain Agents and Model Structure, and (b) training of Target Domain Agents with Policy Reuse. Agents are first trained on source domain tasks to learn task-specific policies and environmental models, which are then stored in the agent pool. For new tasks, the algorithm selects suitable agents from the pool as guide agents to assist with early exploration.

\textbf{PLight: Pre-training of Source Domain Agents and Model Structure}: The left side of Fig.~\ref{fig:overall} represents the situation for the source domain task $k$ . All agents employ a unified architecture comprising: an Encoder extracts intersection observations, a Decoder predicting next-step intersection features, and a Q network evaluating phase action values. During the training process of the source domain tasks, agents explore using the $\epsilon$-greedy method, accumulating experience while learning the corresponding policies and environmental models. The trained agents are then stored in the agent pool to assist in learning other tasks.

\textbf{PRLight: Training of Target Domain Agents and Policy Reuse}: The right module in Fig.~\ref{fig:overall} demonstrates the application of cross-domain knowledge. Our transfer mechanism computes environmental similarity by comparing source agents' predicted feature vectors against target domain observations, enabling optimal guide agent selection. The guide agent outputs an action $a_g$ based on its observation, which interacts with the environment to facilitate behavior transfer. The experiences obtained from interaction with the environment are stored in the experience replay buffer. During training, pre-trained source agents remain fixed during target domain adaptation, ensuring stable knowledge preservation while optimizing transfer efficiency.

\subsection{PLight: Pre-training of Source Domain Agents and Model Structure}
\begin{figure}[h]
    \centering
    \includegraphics[width=\linewidth]{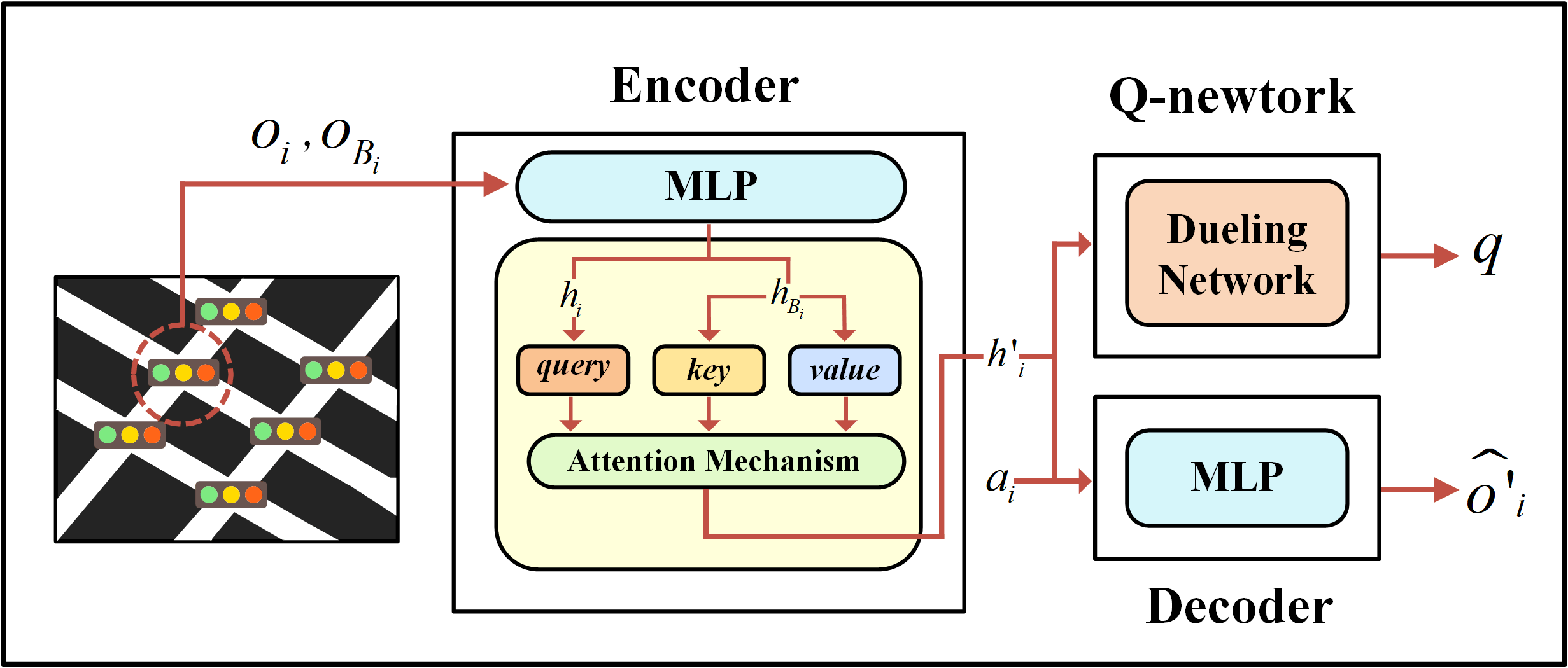}
    \caption{PLight Network Architecture with Depicted Inputs and Outputs}
    \label{fig:PLight}
\end{figure}

In this section, we provide a detailed introduction to the components of the agent model, which are mainly divided into three modules: the Encoder, Decoder, and Q-network. The network structure is illustrated in Fig.~\ref{fig:PLight}. Additionally, we explain the agent's training process in the source domain tasks.

\textbf{Encoder}: The encoder processes the raw observation information of the current intersection to assist the agent in making better decisions by eliminating redundant information and extracting meaningful features. The raw observation information $o_i$ of the intersection $i$ undergoes processing through a multilayer perceptron (MLP) to obtain the corresponding feature vector $h_i$.

\begin{equation}
h_i = Embed(o_i)=\sigma (o_iW_e + b_e),
\label{equ-4-1}
\end{equation}
where $W_e$ and $b_e$ are the network weight matrix and bias vector that need to be learned, and $\sigma$ denotes the ReLU activation function.

Building on this, we use attention mechanisms to integrate neighboring features, providing a more comprehensive and accurate basis for decision-making and prediction.
\[
h'_i = Attention(h_i,h_{B_i}),
\]
$h_{B_i}$ denotes the feature information of the neighbors of agent $i$, which is computed from $o_{B_i}$ according to Eq.~\ref{equ-4-1}, where $o_{B_i}$ denotes the collection of observation information of the neighbors corresponding to agent $i$. Leveraging observations from neighboring agents can enhance collaborative capabilities, and in real-world scenarios, their states are readily observable. The agent's own feature information $h_i$ and the neighbor feature information $h_{B_i}$ are encoded using an attention mechanism into a representation $h'_i$ for subsequent prediction and decision-making.

\textbf{Decoder}: The decoder can be regarded as a forward dynamics model composed of MLPs. It takes as input the features $h'_i$ outputted by the encoder, as well as the actions $a_i$ taken by the agent based on the current observation $o_i$, and predicts the observation of the intersection $\hat{o}'_i$ at the next time step.
\[
\hat{o}'_{i} = Decoder(concatenate(h'_i,a_i)),
\]\

\textbf{Q-network}: The action-value function is composed of a dueling network, which calculates the value of all actions based on the features  $h'_i$ extracted by the encoder. The action with the maximum value is selected for decision-making in the next step.
\[
q = DuelingBlock(h'_i),
\]\

\textbf{Model Training}: The agent in the source domain task starts learning from scratch using the $\epsilon$-greedy method for exploration, continuously interacting with the environment. Once the experience replay buffer reaches a certain size, training begins. The agent employs an Encoder-Decoder structure to learn the environmental state transition model and utilizes an Encoder-Q-network architecture for TSC. During training, decision learning and knowledge storage are optimized simultaneously by minimizing the temporal difference error of the decision module's value function and the mean square error between the prediction module's output and the original observation. The specific loss functions are as follows:
\[
\mathcal{L}_{D}(\varphi, \omega)=\frac{1}{BN}\displaystyle\sum_{b=1}^{B}\displaystyle\sum_{i=1}^{N}\parallel D(o_i,o_{B_i},a_i;\varphi ,\omega ) - o'_{i}\parallel_{2}, 
\]
\[
\mathcal{L}_{Q}(\varphi, \theta)=\frac{1}{BN}\displaystyle\sum_{b=1}^{B}\displaystyle\sum_{i=1}^{N}[{y}_{i}-Q(o'_i,o'_{B_i},a_i;\varphi ,\theta )]^2,
\]

\begin{equation}
\mathcal{L}(\varphi ,\omega ,\theta )=\mathcal{L}_{D}(\varphi, \omega)+\mathcal{L}_{Q}(\varphi, \theta),
\label{equ-4-2}
\end{equation}
where $\varphi$, $\omega$, and $\theta$ represent the network parameters of the Encoder, Decoder, and Q-network, respectively, and $\varphi^-$, $\theta^-$ represent the network parameters of the target networks. $B$ denotes the number of batches of training sampling data. $N$ is the number of agents in the road network. $o_i$ denotes the current observation of agent $i$, while $o_{B_i}$ denotes the observations of its neighbors. $a_i$ denotes the action taken by the current agent. $o'_i$ denotes the observation information of the next time step at intersection $i$. ${y}_{i}=r_i+\gamma \max _{a'}Q(o'_i,o'_{B_i},a';\varphi^- ,\omega^-)$ is the optimization objective of the action value network of the intelligent agent at the intersection $i$. $r_i$ is the reward obtained by the intelligent agent at the intersection $i$ at the current moment. $o'_i$, $o'_{B_i}$ and $a'$ denote the observations and action at the next time step, and $\gamma$ denotes the discount factor.

\begin{algorithm}[h!]
    \caption{PLight-Source Agent Pretraining}
    \label{alg:algorithm-1}
    \textbf{Input:} Randomly initialized model $\mathcal{G}=(\phi, \omega, \theta)$ with Encoder $\phi$, Decoder $\omega$, and Q-network $\theta$ shared by all intersections, and a target model $\mathcal{G}^{-}=\mathcal{G}$.\\
    \textbf{Output:} Learned model $\mathcal{G}=(\phi, \omega, \theta)$. 
    \begin{algorithmic}[1] 
        \STATE Initialize experience replay buffer $\mathcal{D} \leftarrow \emptyset$.
        \FOR {episode = 1, \ldots, E}
        \FOR {$t$ = 1, \ldots, T}
            \STATE For each intersection $i$, selects action $a_{i}$ using $\epsilon$-greedy strategy based on their local and neighbor observations $o_i, o_{B_i}$.
            \STATE Execute joint action $\bm{a_t}=(a_{1},...,a_{n})$ in environment.
            \STATE Receive reward $\bm{r_t}=(r_{1},...,r_{n})$ and next observation $\bm{o'_t}=(o'_{1},...,o'_{n})$, and store transition $(\bm{o_t},\bm{a_t},\bm{r_t},\bm{o'_{t}})$ into $\mathcal{D}$.
            \STATE Sample a batch $B\subset \mathcal{D}$ transactions and update $\phi$, $\omega$, and $\theta$ by Eq.~\ref{equ-4-2}. 
            \STATE Every $k$ steps synchronize target model by $\mathcal{G}^{-}\leftarrow\mathcal{G}$.
        \ENDFOR
        \ENDFOR
    \end{algorithmic}
\end{algorithm}

\subsection{PRLight: Training of Target Domain Agents and Policy Reuse}
Our method accelerates target agent learning through the selection of similarity-weighted guide agents and policy reuse. The core innovation lies in adaptive cross-domain knowledge transfer through environmental similarity measurement.

\begin{figure}[h]
    \centering
    \includegraphics[width=\linewidth]{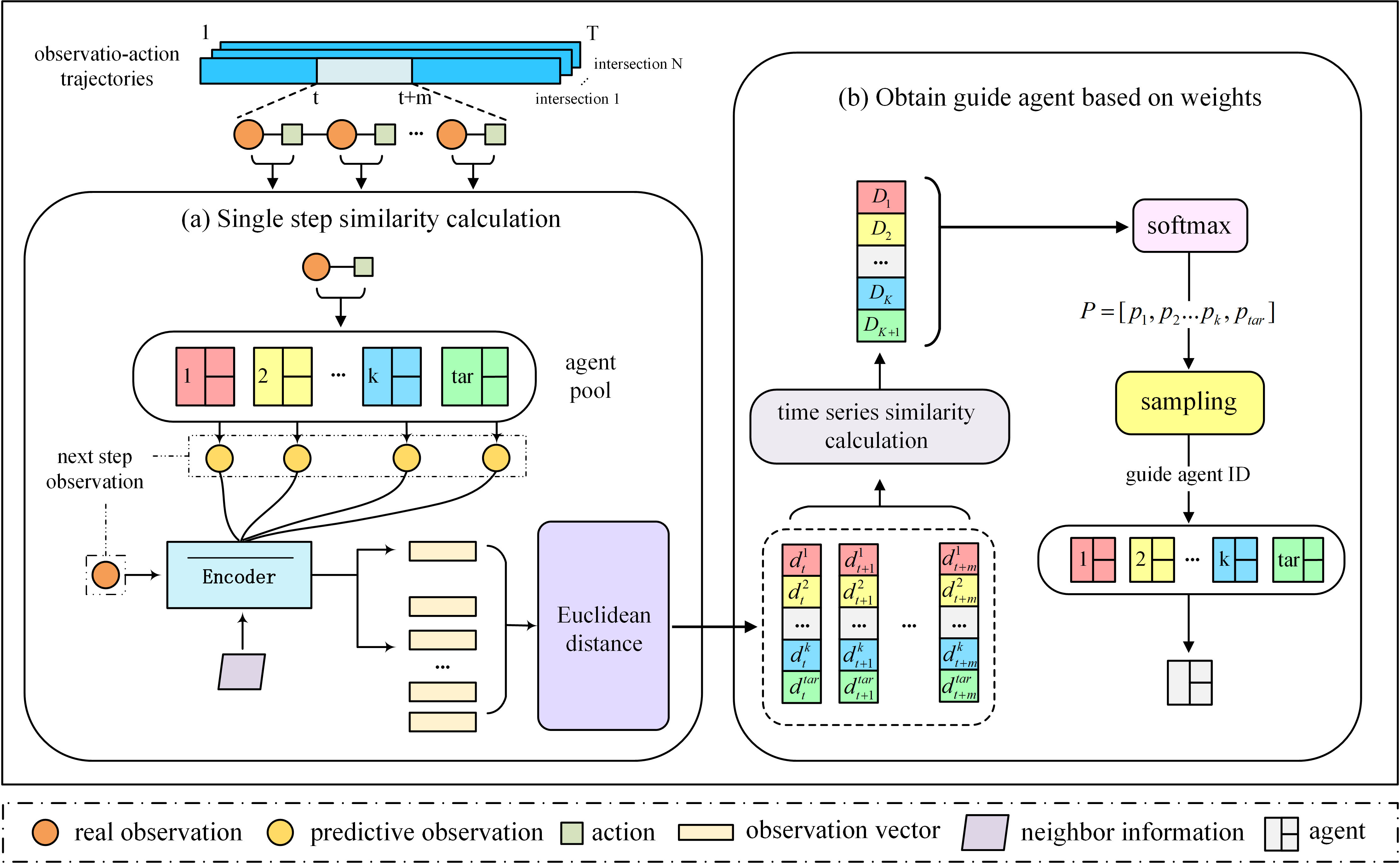}
    \caption{Guide Agent Selection: Single-Step Similarity Computation and Guide Agent Acquisition via Weights}
    \label{fig:selection}
\end{figure}

Traditional random exploration methods suffer from low sample efficiency in unfamiliar traffic environments. To enhance data quality in the experience replay buffer, we implement guided exploration using pre-trained agents. The target agent for each intersection selects optimal guide agents through periodic evaluation of environmental similarity (see Fig.~\ref{fig:selection}), using knowledge from tasks in the source domain with analogous patterns~\cite{van2021multi}. The agent pool contains both fixed source agents and evolving target agents, enabling progressive adaptation to new environments.

\textbf{Single-step similarity calculation:} When the agent interacts with the environment of the target domain, it can obtain a trajectory of observations and actions, the trajectory recorded as a sequence of $m$ time steps representing one cycle. Taking the period from $t\sim t+m$ as an example, at time $t$, the observation feature vector and the action of intersection $i$ are obtained. During the period of $t\sim t+m$, the agents' actions are determined by the guide agent of the previous cycle combined with a small probability of random actions. The reason for adding randomly selected actions is to increase the diversity of experience samples. The observation-action pairs at time $t$ are entered into the agent pool. The agents in the agent pool store the environment models corresponding to the tasks in the source domain. Each model predicts the observation features at time $t+1$ based on the observation-action pairs at time $t$.

To address discrepancies between the predicted and actual observations, we employed an average encoder (obtained by parameter averaging from source encoders) for normalization. Let the normalized true observation feature vector at time $t+1$ be $\mathbf{z'}_{t+1} \in \mathbb{R}^L$ and the corresponding predicted feature vector be $\hat{\mathbf{z'}}_{t+1} \in \mathbb{R}^L$, where \(L\) is the dimension of the characteristic. The Euclidean distance at time \(t\) is then defined as:
\[
d_t = \|\hat{\mathbf{z'}}_{t+1} - \mathbf{z'}_{t+1}\|_2 
= \sqrt{\sum_{l=1}^{L}\left(\hat{z'}_{t+1,l} - z_{t+1,l}\right)^2}.
\]
This distance $d_t$ is used to measure the similarity between the source task and the target task at time $t+1$, where the agent with the smallest distance is more similar to the target task.

\textbf{Obtain guide agent based on weights:} To determine the guide agent, we compute the temporal similarity weight for each agent in the agent pool over the interval \([t, t+m]\). For a given agent \(i\) (with \(i=1,2,\dots,K\) representing the agents of the source domain and \(i=K+1\) representing the target agent), the weight \(D_i\) is calculated so that a smaller cumulative distance (indicating greater similarity) corresponds to a higher weight. It is defined as
\[
D_i = -\sum_{j=t}^{t+m} \lambda^{\,t+m-j}\, d_j^i,
\]
where \(d_j^i = \|\hat{\mathbf{z}}_{j+1}^i - \mathbf{z}_{j+1}^i\|_2\) denotes the Euclidean distance for agent \(i\) at step \(j+1\), and \(\lambda\) is a discount factor that assigns greater importance to more recent time steps. Each agent calculates its similarity weight over this period, resulting in a list of weights \([D_1, D_2, \dots, D_K, D_{K+1}]\) (with \(D_{K+1}\) also denoted by \(D_{\text{tar}}\)). We then transform these weights into a probability distribution via the softmax function:
\begin{equation}
p_k = \frac{\exp\left(D_K\right)}{\displaystyle\sum_{i=1}^{K+1} \exp\left(D_i\right)},
\label{equ-4-3}
\end{equation}
where \(p_k\) is the probability of selecting agent \(k\) as the guide agent. Finally, the guide agent is randomly sampled from the agent pool according to the probability distribution \(\{p_1, p_2, \dots, p_k, p_{tar}\}\). Initial exploration (\(t=0\)) uses random expert sampling until after \(m\) time steps.

\begin{algorithm}[t!]
    \caption{PRLight-Target Agent Training}
    \label{alg:algorithm-2}
    \textbf{Input:} Agent pool $P = \{\mathcal{G}_1, \mathcal{G}_2, \ldots, \mathcal{G}_k, \mathcal{G}_{\text{tar}}\}$, where $\mathcal{G}_i = (\varphi_i, \omega_i, \theta_i)$, period $m$, Experience replay buffer $\mathcal{D}$, Episodes $E$, target network update frequency $k$, Batch size $B$.\\
    \textbf{Output:} Learned model $\mathcal{G}_{\text{tar}}$. 
    \begin{algorithmic}[1] 
        \STATE Initialize: $\mathcal{G}_{\text{tar}} \leftarrow$ randomly select an agent from $P$, $\mathcal{G}^{-}_{\text{tar}} \leftarrow \mathcal{G}_{\text{tar}}$, Experience replay buffer $\mathcal{D} \leftarrow \emptyset$.
        \FOR {episode = 1, \ldots, E}
        \STATE Determine the guide agent $\mathcal{G}_{g}$ for each intersection $i$.
        \FOR {$t$ = 1, \ldots, T}
        \STATE Every $m$ step sample a new guide agent $\mathcal{G}_{g_i} \in P$ for each intersection $i$ by distribution calculated by Eq.~\ref{equ-4-3}.
        \STATE For each intersection $i$, selects action $a^{(G)}_{i}$ according to guide agent $\mathcal{G}_{g_i}$ based on their local and neighbor observations $o_i, o_{B_i}$.
        \STATE Execute joint action $\bm{a_t}=(a^{(G)}_{1},...,a^{(G)}_{n})$ in environment.
        \STATE Receive reward $\bm{r_t}=(r_{1},...,r_{n})$ and next observation $\bm{o'_t}=(o'_{1},...,o'_{n})$, and store transition $(\bm{o_t},\bm{a_t},\bm{r_t},\bm{o'_{t}})$ into $\mathcal{D}$.
        \STATE Sample a batch $B\subset \mathcal{D}$ transactions and update $\mathcal{G}_{\text{tar}}$ by Eq.~\ref{equ-4-2}.
        \STATE Every $k$ steps synchronize target model by $\mathcal{G}^{-}_{\text{tar}} \leftarrow \mathcal{G}_{\text{tar}}$.
        \ENDFOR
        \ENDFOR
    \end{algorithmic}
\end{algorithm}
\textbf{Model Training:} After selecting the guide agent, it makes decisions based on observations and outputs guide action, which interact with the environment to achieve behavior transfer. The experience data from these interactions are stored in the experience replay buffer. Once the replay buffer reaches a certain capacity, the target agent is trained. During training, the source agents in the agent pool remain unchanged, while only the target agent is updated using the same method as the source domain pre-training. The agent pool can include agents trained on different road network tasks, enabling knowledge transfer across various road networks. As the target agent continuously adapts to the new environment, its probability of being sampled gradually increases. Algorithm~\ref{alg:algorithm-2}  outlines the training process of the target agent.

\section{Experiments}
\label{sec:5}

In this section, we design experiments to answer the following questions:

\begin{itemize}
    \item \textbf{RQ1:} Can PRLight maintain stable performance when transferred across different time-of-day (TOD) scenarios in local road networks?
    \item \textbf{RQ2:} Does PRLight exhibit robust performance when generalizing to large-scale, unfamiliar road networks?
    \item \textbf{RQ3:} Does the PLight design choice of sharing a feature extraction module between the policy and environment modules compromise overall policy performance?
\end{itemize}

\subsection{Experimental Setup}

\subsubsection{Datasets}
\begin{figure}[h]
    \centering
    \resizebox{0.9\textwidth}{!}{\includegraphics{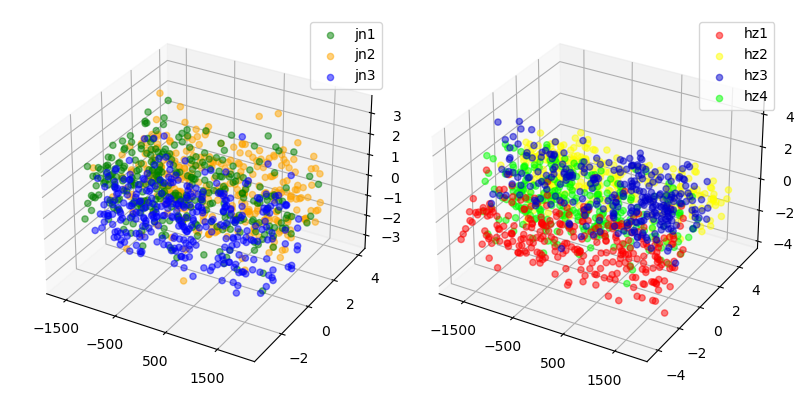}} 
    
    \caption{PCA Analysis of Traffic Flow Data: 3 Datasets from Jinan and 4 Datasets from Hangzhou}
    \label{fig:pca}
\end{figure}

Our experiments utilize CityFlow, an open-source traffic simulator for TSC. Given traffic data as input, the simulator simulates vehicle movements based on predefined environmental settings. Our method generates TSC commands, which the simulator executes before updating the environment state and returning the new state information.

We evaluate our approach on three real-world traffic networks: the large-scale New York road network and synthetically generated datasets for Jinan and Hangzhou. The New York dataset reflects actual traffic conditions, while the Jinan and Hangzhou datasets simulate different time-of-day (TOD) scenarios, each covering an hour of traffic flow. The Jinan dataset includes three traffic flow scenarios, and the Hangzhou dataset contains four. The road networks comprise 3×4 and 4×4 intersections, respectively. Traffic flow data is generated by the frequency of entering the road network and the probability of vehicle direction. Detailed settings are provided in Table~\ref{tab:2}.

\begin{table}[h]
    \centering
    \caption{Traffic Flow Dataset Configuration for Jinan and Hangzhou Road Networks. Taking the \textbf{hz1} dataset as an example, the simulation spans 1 hour, divided into three phases with varying vehicle entry frequencies. In the first 20 minutes, vehicles enter the network approximately every 7 seconds from different entry points, following probabilistic routes (straight, right, or left) until exiting. A total of 6,684 vehicles entered the Hangzhou network during this period.}
    \label{tab:2}
    
    \begin{minipage}{\textwidth}
        \centering
        \captionof{subtable}{Jinan dataset traffic distribution settings}
        \label{tab:subtable1}
        \begin{tabular}{c|c|c|c|cc|c}
        \hline
            Flow & straight  & right   & left   & \multicolumn{2}{c|}{frequency}  & vehicles \\ 
        \hline
            jn1  & 0.3 & 0.3 & 0.4 & \multicolumn{1}{c|}{9}  &  5 & 7831  \\
            jn2  & 0.4 & 0.4 & 0.2 & \multicolumn{1}{c|}{5.5}  & 5.5 & 9172  \\ 
            jn3  & 0.5 & 0.3 & 0.2 & \multicolumn{1}{c|}{8}  &  5 & 8186  \\ 
        \hline
        \end{tabular}
    \end{minipage}
    
    \vspace{1em}
    
    \begin{minipage}{\textwidth}
        \centering
        \captionof{subtable}{Hangzhou dataset traffic distribution settings}
        \label{tab:subtable2}
        \begin{tabular}{c|c|c|c|ccc|c}
        \hline
            Flow & straight  & right   & left   & \multicolumn{3}{c|}{frequency}  & vehicles \\ 
        \hline
            hz1  & 0.6 & 0.15 & 0.25 & \multicolumn{1}{c|}{7}  & \multicolumn{1}{c|}{10.2} & 9.3 & 6684  \\ 
            hz2  & 0.1 & 0.7  & 0.2  & \multicolumn{1}{c|}{10} & \multicolumn{1}{c|}{11}   & 4   & 8444  \\ 
            hz3  & 0.2 & 0.3  & 0.5  & \multicolumn{1}{c|}{4}  & \multicolumn{1}{c|}{10}   & 10  & 8433  \\ 
            hz4  & 0.3 & 0.4  & 0.3  & \multicolumn{1}{c|}{8}  & \multicolumn{1}{c|}{5}    & 4   & 11012 \\ 
        \hline
        \end{tabular}
    \end{minipage}
    
\end{table}
 
To visually distinguish between the datasets within the network, we performed Principal Component Analysis (PCA) on the traffic flow files for visualization. The feature vectors for vehicles entering the network are made up of two aspects: the distribution of vehicle directions and the time of entry into the network. We encoded the direction one hot (\eg., straight as [1,0,0]), recording the vector $p_t$ each time a vehicle moves to the next lane. The feature vector for each turn is multiplied by a corresponding discount factor to differentiate between routes. The vectors after each directional change are summed to obtain the feature vector representing the vehicle's route direction distribution, denoted as:
\[
v_{turn}=\displaystyle\sum_{i=0}^{n}\sigma^{i}p_{i},
\]
In the above equation, $n$ denotes the total number of turns a vehicle makes on its route, and $p_i$ denotes the eigenvector on the $i$-th turn. The direction feature vector, combined with the time of entry into the network, constitutes the travel feature vector $V_{travel} = [v_{turn}, t_{start}]$. By mapping the feature vectors of the traffic flow to a three-dimensional space, the distribution information of the data can be visualized, as shown in the Fig.~\ref{fig:pca}.

\subsubsection{Compared Methods}
\begin{itemize}
    \item \textbf{CoLight~\cite{wei2019colight}}: This method utilizes a graph attention network to enhance the connections between the current intersection and its neighboring intersections, enabling coordinated control.
    \item \textbf{AttendLight~\cite{oroojlooy2020attendlight}}: This method employs an attention mechanism to construct phase features and predict the probability of phase changes.
    \item \textbf{MPLight~\cite{zhang2022expression}}: This method uses FRAP as the base model and incorporates pressure into the design of states and rewards.
    \item \textbf{AttentionLight~\cite{zhang2023leveraging}}: This method effectively represents and designs queue lengths in the state and uses self-attention to learn phase correlations, achieving TSC.
\end{itemize}

\subsubsection{Evaluation Metric}
Based on current research, we use average travel time, throughput, and queue length to assess the performance of TSC algorithms. The definitions are as follows:

\begin{itemize}
    \item Travel Time($m_{tt}$): The average travel time of all vehicles.
    \item Throughput($m_{th}$): The total number of vehicles reaching their destinations.
    \item Queue Length($m_q$): The average queue length at each intersection.
\end{itemize}

In addition, to evaluate the efficiency of the learning process, we introduce two common metrics in transfer learning, both based on average travel time:

\begin{itemize}
    \item Accumulated Rewards($ar$): The area under the agent’s reward curve reflects the algorithm’s overall performance throughout the learning process. Our experiment uses the area enclosed by the average travel time curve in the training cycle as the evaluation $ar$. The magnitude of this metric is on the order of $10^5$.
    \item Performance with Fixed Training Epochs($pe$): The algorithm's performance after a specified number of training epochs. We use the average travel time after 3 training epochs as the evaluation metric for $pe$.
\end{itemize}

\begin{table*}[ht!]
\centering
\setlength\tabcolsep{3pt}
\caption{Execution performance comparison. (Top values are in \textbf{bold}, $\uparrow$ indicate that larger values are better and $\downarrow$ indicate that smaller values are better)}
\label{table:Performance}
\resizebox{1\textwidth}{!}{ 
    \begin{tabular}{c|c|cccccc}
    \hline
    \multicolumn{2}{c|}{Metrics}  & CoLight & MPLight  & AttendLight & AttentionLight  & PLight   & PRLight \\ 
    \hline
    \multirow{5}{*}{jn3}
                           &${m_{tt}} \downarrow$ &498.36 &441.81 &416.49 &\textbf{374.97}   &455.62   &375.67 \\
                           &${m_{th}} \uparrow$  &7055 &7087 &7069 &\textbf{7351}   &6894   &7346  \\
                           &${m_{q}} \downarrow$ &57.09 &46.95 &40.93 &28.44   &46.19   &\textbf{25.32} \\
                           &${ar} \downarrow$  &2.17556 &1.49837 &1.43734 &1.29838   &2.04319   &\textbf{1.10052}\\
                           &${pe} \downarrow$  &1292.97 &864.59 &540.25 &575.54   &1328.57   &\textbf{381.96}  \\
    \hline
    \multirow{5}{*}{hz3}
                           &${m_{tt}}$ $ \downarrow $ &625.39 &576.13 &570.62 &497.81   &646.92  &\textbf{473.71}  \\
                           &${m_{th}}$ $ \uparrow $ &8131 &8207 &8157 &8173   &8198   &\textbf{8238}  \\
                           &${m_{q}}$ $ \downarrow $ &35.24 &38.02 &41.56 &26.29   &33.99   &\textbf{22.17}  \\
                           &${ar}$ $ \downarrow $ &2.55912 &2.196235 &1.93544 &1.75715   &3.03232   &\textbf{1.40221}\\
                           &${pe} \downarrow$  &1735.63 &1090.91 &757.53 &1035.09   &1728.95   &\textbf{487.61}  \\
    \hline
    \multirow{5}{*}{newyork}
                           &${m_{tt}}$ $ \downarrow $ &\textbf{1043.84} &1601.04 &1491.86 &1155.00   &1261.43   &1054.69\\
                           &${m_{th}}$ $ \uparrow $ &\textbf{8480} &2538 &4729 &7711   &6394   &7818\\
                           &${m_{q}}$ $ \downarrow $ &9.12 &17.79 &14.61 &10.69  &12.36   &\textbf{8.19}\\
                           &${ar}$ $ \downarrow $ &3.73232  &4.47499  &4.16902 &3.50759   &4.00278   &\textbf{3.23941}\\
                           &${pe}$ $ \downarrow $ &1718.73 &1454.46 &1423.26 &1269.86   &1707.27   &\textbf{1188.67}\\
    \hline
    \end{tabular}
}
\end{table*}

\subsection{Experiment 1: Within-Network Transfer Experiments}
We set up two sets of network transmission experiments. To better design transfer experiments and explore the effectiveness of our method, reflecting the differences in traffic flow datasets. We propose two concepts based on the number of vehicles and traffic flow distribution in the traffic flow dataset.

\begin{figure}[ht]
    \centering
    \resizebox{0.9\textwidth}{!}{\includegraphics{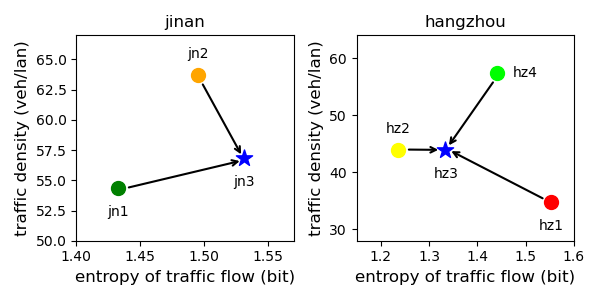}} 
    \caption{Two groups of transfer experiment settings. In the Jinan network, jn1 and jn2 serve as source domains, with jn3 as the target domain, whose distribution lies outside the source domains. In the Hangzhou network, hz1, hz2, and hz4 are source domains, while hz3 is the target domain, whose distribution remains within the source domains.}
    \label{fig:distribution}
\end{figure}

\textbf{Traffic Flow Density}. The number of vehicles entering an intersection is a critical factor in TSC. We define traffic flow density as the number of vehicles per lane per hour, which represents traffic volume. The higher density of traffic flow indicates more vehicles on a road segment, leading to increased congestion, while the lower density suggests fewer vehicles and smoother traffic conditions. The formula for traffic flow density is:

\[
E_{\rho }=\frac{N{veh}}{N{lan}},
\]
where $N{veh}$ denotes the total number of vehicles passing through all intersections in one hour, and $N{lan}$ denotes the number of lanes.

\textbf{Traffic Flow Distribution Entropy}. The distribution of the probability of steering in road networks varies between datasets, reflecting the complexity of traffic conditions. We quantify this through entropy $H$, \ie, an information-theoretic measure of the uncertainty of the event. Higher entropy implies more potential vehicle routes and complex traffic scenarios. To address entropy ambiguity in conventional probability distributions (\eg, [0.2, 0.3, 0.5] vs. [0.3, 0.2, 0.5]), we propose the Cumulative Normalization Transformation (CNT). This method generates discriminative distributions through cumulative normalization, enabling effective differentiation of entropy characteristics for analyzing vehicle steering probabilities (straight/left/right turns). The formula for the entropy of the traffic flow distribution is:
\[
H = -\displaystyle\sum_{i=1}^{3}(\frac{1}{S}\displaystyle\sum_{j=1}^{i}p_j·\log_{2}{\frac{1}{S}\displaystyle\sum_{j=1}^{i}p_j}),
\]
where $S=p_1+p_2+p_3$ is a normalization factor. $p_1$, $p_2$, and $p_3$ denote the probabilities that vehicles continue straight, turn right, and turn left at intersections in the road network, respectively.

\begin{figure}[ht]
    \centering
    \includegraphics[width=\linewidth]{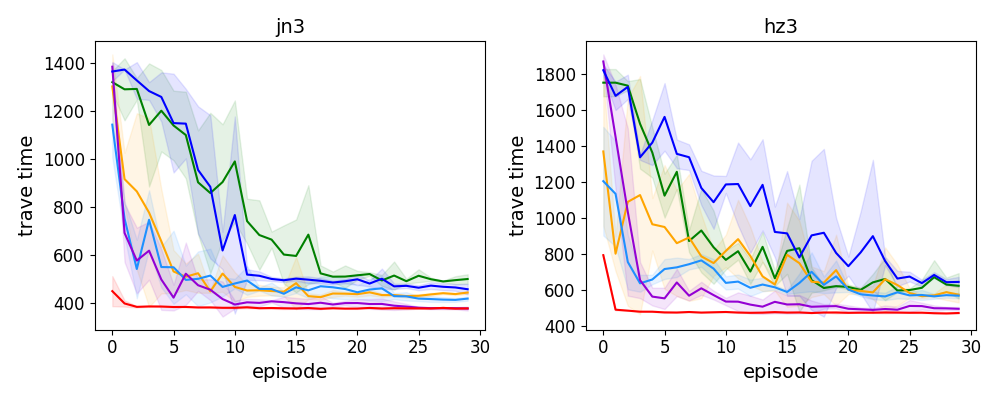} 
    \caption{Travel time performance comparison of RL methods (CoLight `\textbf{{\color[HTML]{008000}---}}', MPLight `\textbf{{\color[HTML]{FFA500}{---}}}', AttendLight `\textbf{{\color[HTML]{1E90FF}{---}}}', AttentionLight `\textbf{{\color[HTML]{9400D3}{---}}}', PLight `\textbf{{\color[HTML]{0000FF}{---}}}' and PRLight `\textbf{{\color[HTML]{FF0000}{---}}}') in Within-Network Transfer Experiments: Testing Curves on jn3 and hz3 Datasets over 30 Episodes}
    \label{fig:in}
\end{figure}

We conducted two experiments to assess the influence of source policies on target policies and to verify the robustness of the transfer algorithm. In these experiments, the density of the traffic flow is plotted on the vertical axis and the distribution entropy on the horizontal axis (see Fig.~\ref{fig:distribution}). Fig.~\ref{fig:in} illustrates the test curves obtained from evaluating various comparison methods alongside our proposed approach, using the travel-time metric over 30 episodes on both the jn3 and hz3 datasets. Notably, the general pre-trained PLight model consistently underperformed relative to several mainstream TSC methods. This discrepancy is probably due to the relatively simplistic architecture and design of PLight. In stark contrast, our transfer learning method, PRLight, not only achieved superior performance but also demonstrated markedly faster convergence, requiring only 5 episodes to reach a stable performance level.  This shows that PRLight effectively uses knowledge in the source domain to accelerate the learning of target tasks. Furthermore, the test curves in Fig.~\ref{fig:in} reveal that PRLight exhibits significantly lower variance, thereby demonstrating its robustness and adaptability in dynamic traffic environments.

\subsection{Experiment 2: Cross-Network Transfer Experiments}
\begin{figure}[ht]
    \centering
    \includegraphics[width=\linewidth]{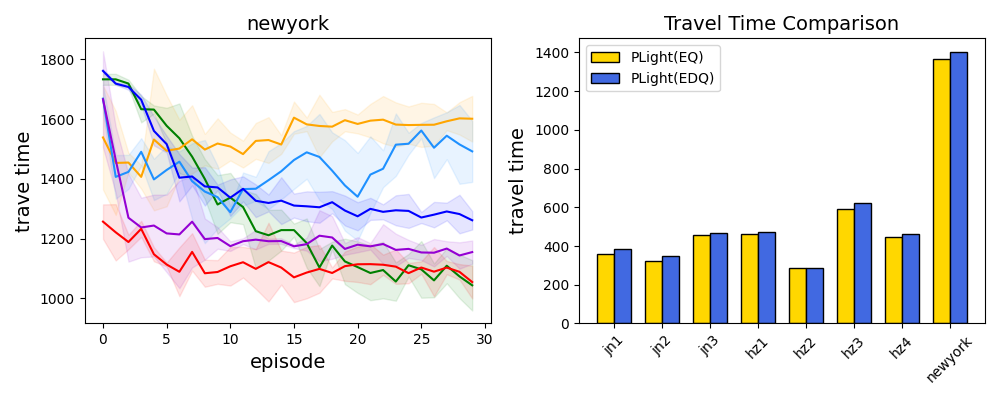} 
    \caption{Travel time performance comparison of RL methods (CoLight `\textbf{{\color[HTML]{008000}---}}', MPLight `\textbf{{\color[HTML]{FFA500}{---}}}', AttendLight `\textbf{{\color[HTML]{1E90FF}{---}}}', AttentionLight `\textbf{{\color[HTML]{9400D3}{---}}}', PLight `\textbf{{\color[HTML]{0000FF}{---}}}' and PRLight `\textbf{{\color[HTML]{FF0000}{---}}}') in Cross-Network Transfer (New York) and Environmental Model Impact PLight(EDQ/EQ) over 30 Episodes}
    \label{fig:cross2melt3}
\end{figure}

The New York road network is a large-scale grid with dimensions 27×8, comprising 196 intersections. To evaluate the generalization performance of our method in an unfamiliar and complex environment, we transferred the policies originally developed for the Jinan and Hangzhou road networks to the New York network. In this cross-network transfer experiment, the policies obtained from the jn3 and hz3 target domains during the within-network experiments were employed as expert policies, thereby demonstrating the reusability of our approach. The corresponding results are presented in Fig.~\ref{fig:cross2melt3}.

CoLight, enhanced by a graph attention network, excels in large-scale road networks due to its adaptability. In contrast, MPLight and AttendLight fail to converge, likely because they are ill-suited to the complexities of large-scale networks and cannot develop effective policies. PRLight outperforms most methods in Experiment 1 but falls short of CoLight in some metrics in Experiment 2. This may be due to the use of homogeneous agents for transfer learning in Experiment 1. In the New York experiment, agents trained on the Jinan and Hangzhou networks were transferred without adaptation, limiting their exploration and resulting in suboptimal policies. Despite these challenges, PRLight's strengths remain intact. It surpasses most methods in learning speed and efficiency, and agents trained in small-scale synthetic datasets can effectively aid learning in large-scale tasks. Although the New York experiment shows higher variance than Experiment 1, PRLight still demonstrates superior stability and adaptability compared to other approaches.

\subsection{Experiment 3: Investigating the Impact of Environmental Models on policy Performance}

To further assess whether learning an environmental model influences policy performance, we conducted an experiment comparing two network variants. The complete network structure PLight(EDQ) incorporates a Decoder environmental model, while the alternative variant PLight(EQ) operates without the Decoder component. Both models were trained under identical conditions and evaluated for overall policy performance. As shown in Fig.~\ref{fig:cross2melt3}, the experimental results revealed that PLight(EQ) achieved marginally better performance metrics compared to PLight (EDQ), although the difference did not reach statistical significance. No substantial disparities in training speed were observed between the two architectures. This suggests that the inclusion of the Decoder module during training does not substantially impact the learning process of the policy component, as both variants demonstrated comparable learning outcomes.

\section{Conclusion}
\label{sec:6}

To address the challenges of poor generalization, low sampling efficiency, and slow training speed in deep reinforcement learning-based TSC methods for new environments, we applied transfer learning techniques. Our PLight method, which constructs an environmental state transition model while learning agent decision-making policies, and the PRLight method, designed for policy reuse, demonstrate strong generalization capabilities and learning efficiency. Experimental results validate the effectiveness of our approach, showing its superiority in stability and exploration cost over popular methods.

In future work, exploring strategies for the implementation of our method in complex road networks, such as those with high density intersections and diverse transport facilities, is a promising research direction. Also, since we have reduced the training scale, further studying how to reduce the inference scale to enhance the method's applicability and efficiency is worth exploring. These two directions will help improve the generalization of our method and its practical value.

\section*{Acknowledgments}
This work was supported by the National Natural Science Foundation of China under Grants 62376048 and 62406050, and the Shandong Provincial Natural Science Foundation, China, under Grant ZR2022LZH002.

\bibliographystyle{elsarticle-num-names}
\bibliography{main}

\end{document}